\documentclass[authoryear,preprint,12pt]{elsarticle}
\pdfoutput=1 
\makeatletter
\def\ps@pprintTitle{%
	\let\@oddhead\@empty
	\let\@evenhead\@empty
	\let\@oddfoot\@empty
	\let\@evenfoot\@oddfoot
}
\makeatother

{
\usepackage{array} 
\usepackage{textcomp} 
\usepackage[breaklinks]{hyperref} 
\usepackage[table,xcdraw]{xcolor} 
\graphicspath{ {images/} } 
\usepackage[none]{hyphenat} 
\usepackage{siunitx} 
\usepackage{amssymb} 

\usepackage{subfig} 
\usepackage{amsmath} 
\usepackage{mhchem}

\journal{Icarus}

\begin{document} \sloppy 

\begin{frontmatter}

\title{ Bistatic Radar Observations of Near-Earth Asteroid (163899) 2003 SD220 from the Southern Hemisphere}

\author[label1]{Shinji Horiuchi} \ead{shoriuchi@cdscc.nasa.gov} 
\author[label1,label2]{Blake Molyneux} 
\author[label3]{Jamie B. Stevens} 
\author[label1]{Graham Baines} 
\author[label4]{Craig Benson} 
\author[label4]{Zohair Abu-Shaban} 
\author[label5]{Jon D. Giorgini} 
\author[label5]{Lance A.M. Benner} 
\author[label5]{Shantanu P. Naidu}
\author[label3]{Chris J. Phillips} 
\author[label3]{Philip G. Edwards} 
\author[label1]{Ed Kruzins} 
\author[label6]{Nick J.S. Stacy} 
\author[label5]{Martin A. Slade} 
\author[label3]{John E. Reynolds} 
\author[label5]{Joseph Lazio}

\address[label1]{CSIRO Astronomy and Space Science, Canberra Deep Space Communication Complex, Tidbinbilla, ACT, Australia} 

\address[label2]{School of Engineering, Australian National University, Canberra, ACT, Australia} 

\address[label3]{CSIRO Astronomy and Space Science, Australian Telescope National Facility, Epping, New South Wales, Australia} 

\address[label4]{School of Engineering and Information Technology, University of New South Wales, Canberra, ACT, Australia} 

\address[label5]{Jet Propulsion Laboratory, California Institute of Technology, Pasadena, CA, USA}

\address[label6]{Defence Science and Technology Group, Adelaide, South Australia, Australia}

\begin{abstract} We report results of Canberra-ATCA Doppler-only continuous wave (CW) radar observations of near-Earth asteroid (163899) 2003 SD220 at a receiving frequency of 7159 MHz (4.19 cm) on 2018 December 20, 21, and 22 during its close approach within 0.019 au (7.4 lunar distances).  Echo power spectra provide evidence that the shape is significantly elongated, asymmetric, and has at least one relatively large concavity.  An average spectrum per track yields an OC (opposite sense of circular polarisation)  radar cross section of 0.39, 0.27, and 0.25 km$^{2}$, respectively, with an uncertainty of 35 \%. 
Variations by roughly a factor of two
in the limb-to-limb bandwidth over the three days indicate rotation of an elongated object. We obtain a circular polarization ratio of 0.21 $\pm$ 0.07 that is 
consistent with, but somewhat lower than, the average among other S-class near-Earth asteroids observed by radar.

\end{abstract}

\begin{keyword}

Asteroid \sep NEAs \sep 2003 SD220 \sep radar \sep Deep Space Network 
\sep  ATCA

\end{keyword}

\end{frontmatter} 


\section{Introduction} \label{section1}

Near-Earth asteroid (163899) 2003 SD220 is an Aten-type discovered on 29 September 2003 by the Lowell Observatory Near-Earth Object Search (LONEOS).  The absolute magnitude of 2003 SD220 is $H = $17.2.  NEOWISE thermal infrared observations suggested an effective diameter of 0.8 $\pm$ 0.2 km (Nugent et al. 2016).  Using optical and near-infrared spectroscopy observations, DeMeo et al. (2014), Perna et al. (2016), and Rivera-Valentin et al. (2019) classified this object as an S/Sr-type, which are 
indicative of a siliceous (i.e. stony) mineralogical composition and thought to be analogous to the H ordinary chondrite meteorites (e.g. Gaffey and Gilbert 1998).
2003 SD220 appears on the NASA  Near-Earth Object Human Space Flight Accessible Targets of Study (NHATS) list (https://cneos.jpl.nasa.gov/nhats/), and has been considered as a possible mission target (Kawakatsu et al. 2009, P. W. Chodas, pers. comm.), so detailed characterization is important to facilitate possible future human or robotic exploration.  

Initial radar results from Arecibo and Goldstone in December 2015 revealed that 2003 SD220 is elongated with a long axis exceeding 2 km and an extremely slow rotation period of roughly 12 days (Rivera-Valentin et al. 2019). A period of 285 
$\pm$ 5 hours was obtained with lightcurve data and hints of non-principal axis rotation were identified (Warner 2016).  

2003 SD220 recently made an extremely close approach to Earth on 2018 December 22, when it passed by at 0.0189 au (7.4 lunar distances), the closest it will come to the Earth until 2070; the next close approach will be at 0.0363 au (14.2 lunar distances) on 2021 December 17. The asteroid's close approach, size, and very slow rotation produced extremely strong radar signal-to-noise ratios (SNRs) in 2018.  2003 SD220 approaches Earth closely every three years between 2015-2027 and radar observations at each encounter provide an opportunity for 
precise orbit monitoring,
imaging and detailed 3-D shape and spin state estimation.  
Multi-station radar and visible-infrared spectroscopic observations were coordinated between Arecibo Observatory, Goldstone, the Green Bank Telescope (GBT), the Quasar VLBI network, and the NASA Infrared Telescope Facility (IRTF) from 
the northern hemisphere (Rivera-Valentin et al. 2019, Bondarenko et al. 2019), and the Deep Space Network Stations in Tidbinbilla, near Canberra, and the Australian Telescope Compact Array (ATCA) in Narrabri from the southern hemisphere. 
Rivera-Valentin et al. (2019) obtained high resolution range-Doppler images with bistatic radar observations using Arecibo-GBT and Goldstone-GBT, revealing slowly rotating and very elongated structure of 2003 SD220 as "batata-like" (or sweet-potato-shaped). They also conducted infrared spectroscopy and concluded that 2003 SD220 is a S-complex asteroid.
International multi-station CW (continuous wave) Doppler-only observations were conducted using Goldstone to transmit and the 64-m Sardinia Radio Telescope (SRT) to receive the echoes from 2003 SD220, which Bondarenko et al. (2019) eavesdropped using the 32 m Zelenchukskaya and Svetloe radio telescopes in Russia and confirmed the elongated nature of this asteroid.

Here we present results of the southern hemisphere radar observations from 2018. The radar observations from Australia occurred at times when radar observations were not scheduled at Arecibo and Goldstone because the asteroid was below the horizon or due to scheduling conflicts. Consequently, our observations provide independent information on the physical properties at orientations not observed at other facilities.  The radar echoes were very strong and produced data that will provide valuable contributions for 3-D shape estimation using radar images obtained at Arecibo, Goldstone, and Green Bank.


\section{Observations} \label{Section2}

 We observed 2003 SD220 in Australia, on 2018 December 21, 22, and 23, dates that straddled its closest approach to Earth. 
 We employed a bistatic approach, for each day using one of the three Deep Space Network (DSN) C-band (7259 MHz, 4.1 cm) 20 kW power transmitters at the three Deep Space Stations, DSS-43, 35, and 36, 
of the Canberra Deep Space Communication Complex (CDSCC) in Tidbinbilla with reception at the Australian Telescope Compact Array (ATCA) in Narrabri, which is 566 km to the north. 
 Our observation and data reduction techniques are similar to those described by Benson et al. (2017a), Benson et al. (2017b),  and Abu-Shaban et al. (2018) for previous NEA radar observations with the southern hemisphere planetary radar system that utilised the Parkes Radio Telescope to receive echoes at a transmitter frequency of 2.1 GHz. For 2003 SD220, we used the ATCA to receive echoes
 centred on 7159.45 MHz,
 a frequency that could not be received at Parkes. 

 Table 1 summarises the observations. The ATCA conducted observations in the 1.5D East-West array with a separation of each antenna toward the West 
from the first antenna, CA01, of 107 m for CA02, 582 m for CA03, 1224 m for CA04, 1439 m for CA05, and 4439 m for CA06. 
 Both a phased array (tied-array mode) of up to 5 antennas (excluding CA06) referenced to CA03, and a single antenna (CA03) 
 were recorded simultaneously.

 To track the asteroid on all three dates we used JPL Horizons orbit solution 94, which is based on 
 444 ground-based visual-spectrum position measurements, along with 59 infrared position measurements relative to the WISE spacecraft, 
 and 6 delay and 4 Doppler measurements from prior radar observations at Arecibo and Goldstone in 2015 and 2018. We transmitted a right-circularly polarised (RCP) electromagnetic wave that reflects off the target asteroid in both the same (SC) and opposite (OC) senses of circular polarisation as the transmitted wave.  Using the predicted trajectory of the asteroid, the transmitted frequency was continuously adjusted so that the received echo frequency 
at a constant reference frequency at ATCA, centered on 7159.45 MHz.
 
 The Long Baseline Array Data Recorders (LBADR, Phillips et al. 2009) were used to record a 4 MHz bandwidth in dual polarization. The system noise temperature at ATCA varied between 17 K to 25 K, depending  primarily on the weather and target elevation.   \

During the observations the array was moved to a nearby 
Active Galactic Nuclei (AGN) radio source
every ~90 minutes to phase up the array. On Dec 20-21 we used J1608+1029, on Dec 21-22 we used J1707+0148, and on Dec 22-23 we used J1743-0350.

\begin{table}[h!] 
\begin{center} 
\footnotesize 
\setlength{\tabcolsep}{6pt} 
\begin{tabular}{lccccccc} 
\hline 
Date & Start-Stop UTC & R.A. & Dec. & Distance & Transmit  ant. & G$_{tx}$ & Tsys \\ 
(yy-mm-dd)& (number of spectra) & ($^{\circ}$) & ($^{\circ}$) & (au) & (diameter) & (dB) & (K) \\ 
\hline 
2018 Dec 20-21  & 23:49:12-03:35:51 (218) & 16.7 & +17.3 & 0.0193 & DSS-43 (70 m) & 73.23 & 25  \\
2018 Dec 21-22 & 21:01:48-03:25:25 (361) & 17.1 & +8.2 & 0.0189 & DSS-35 (34 m) & 66.98 & 23  \\
2018 Dec 22-23 & 22:44:48-02:40:18 (210) & 17.5 & -0.3 & 0.0192 & DSS-36 (34 m) & 66.98 & 17  \\
\hline 
\end{tabular} 
\end{center} 
\caption{
List of all experiments involving transmission with a Doppler-shift compensating time-varying frequency that produced reception at a constant reference frequency at ATCA, centered on 7159.45 MHz, in both OC and SC polarizations.  
On each line we give the observing date; the starting and ending receive times (number of 60 second integrated spectra to average, excluding time required to stop observations for ATCA phase calibration); average right ascension (R.A.) and declination (Dec.) in 
ICRF 
and average distance from the Earth; name of the Deep Space Station (and its diameter) used to transmit, and its transmitter gain G$_{tx}$; average ATCA receiver system noise temperature Tsys.} 
\label{table: 1} 
\end{table}

Given the standard deviation of the receiver noise as a function of integration time, frequency resolution, and system noise temperature, the radar cross section  $\sigma$ can be computed from integrating the echo power spectra using the radar equation (e.g. Ostro 1993): 

\begin{equation} 
\label{equation_new}
P_{rx} = \frac{P_{tx} G_{tx} G_{rx}\lambda^{2}\sigma}{(4\pi)^{3} R^{4}} 
 \end{equation} 
 
where $P_{tx}$ is the transmitter power (in watts), $G_{tx}$ and $G_{rx}$ are the gain of the transmitting and receiving antennas, $\lambda$ is the radar wavelength, $\sigma$ is the radar cross-section of the target, and $R$ is the distance between the target and the reference point of the antenna. $P_{tx}$ is 20 kW on average for all the transmitting stations here. $G_{tx}$ value for each Deep Space Station used is shown in Table 1. 
We adopt the ATCA receiver gain $G_{rx} $ = 61.59 dB, the average value measured via holography in 2014 for all antennas except for CA03, 
which were the same within 0.6 dB.

The standard deviation of noise of the echo power spectrum $\Delta P_{noise}$ is given by 
\begin{equation} 
\label{equiation2} 
\Delta P_{noise} = \frac{k T_{sys} \Delta f} {( \Delta t \Delta f)^{\frac{1}{2}}} 
\end{equation} 

where $k$ is Boltzman's constant, $T_{sys}$ is the receiver system noise temperature, $\Delta f$ is frequency resolution, and $\Delta t$ is total integration time of the received signal. 

Using the equations (1) and (2) we estimate the cross section by integrating  under the points containing echo power SNR (= $P/\Delta P_{noise}$ ) using calibrated data.  The uncertainties in  $\sigma$ are dominated by systematic pointing and calibration errors that are typically between 20 and 50 \%; based on our experience with receiving at the ATCA, we adopt a 35 \% error.

The Doppler broadening or bandwidth $B$ of the echo is defined as 

\begin{equation} 
\label{equiation3} 
B = \frac{4 \pi D(\phi) \cos \delta} {\lambda P} 
\end{equation} 

where $P$ is the rotation period, 
$D(\phi)$ is the diameter at rotation phase $\phi$,
and $\delta$ is the angle between the radar line-of-sight and the asteroid's apparent equator (i.e., the subradar latitude). If $P$ is known, then measuring $B$ and setting cos $\delta$ = 1, 
corresponding to an equatorial view or sub-radar latitude of zero, 
places a lower bound on the asteroid's maximum pole-on breadth $D_{max}$. Expressing $B$ in Hz, $D$ in kilometers, and $P$ in hours gives  $B$(Hz) = 
83.3 $D$(km)cos $\delta$/$P$(hr) at $\lambda$ = 4.189 cm.

\section{Results} \label{Section3}

For both polarizations, the baseband data were transformed into the frequency domain, using a coherent processing interval of 60 seconds to yield a frequency resolution of 0.0167 Hz that is sufficient to resolve the echo's bandwidth. The individual spectra were then accumulated incoherently over a sufficient duration to detect the echo from the asteroid. 

The ATCA operated as a tied array and as a single antenna. We detected strong echoes during all three tracks with both the tied array and single antenna. Given the strong detections, with daily average SNRs $>$ 50 even for the single antenna, we decided to focus on only the single antenna data for further analysis to avoid decoherence with the tied array due to time-varying atmospheric instability. 
Because of uncertainty in the ATCA reference position astrometry was not attempted for this data.
Figure 1 shows integrated echo power spectra for each day obtained with the single antenna. Figures 2, 3, and 4 show echo power spectra for every 60 minute interval of each track at a resolution of 0.0167 Hz for Dec. 20-21, Dec. 21-22, and Dec. 22-23, respectively.

\begin{figure}[!ht] 
\centering 
\includegraphics[width=0.88\textwidth]{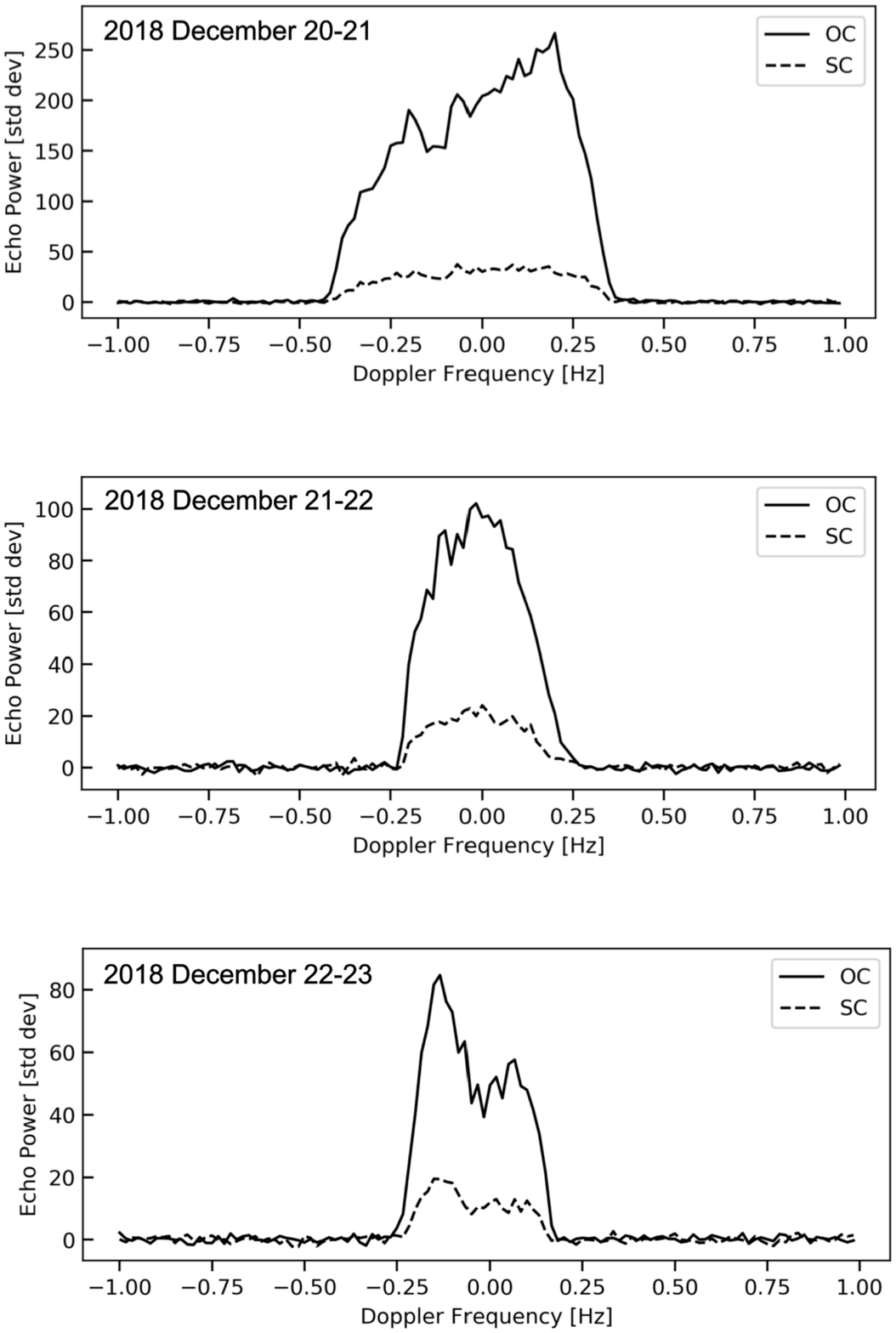} 
\caption{
Echo power spectra for 2003 SD220, averaging all data from each day, shown at a frequency resolution of 0.0167 Hz. The 0 point of the x-axis is 7159.45 MHz. 
Transmitter antenna used is DSS-43 (top, 70m), DSS-35 (middle, 34m), and DSS-36 (bottom, 34m).
} 
\label{fig: correction} 
\end{figure}

Table 2 lists echo bandwidths and disk-integrated properties such as radar cross section for each track.   \

\begin{table}[!h] 
\begin{center} 
\small 
\setlength{\tabcolsep}{7pt} 
\begin{tabular}{lcccccc} 
\hline 
Date  & $\Delta \phi$ (deg) & OC SNR & $B$(Hz) & $\sigma_{OC}$(km$^2$) & $D_{max}$(km) & SC/OC \\ 
\hline 
2018 Dec 20-21 & 0 - 5 & 267 & 0.79   & 0.39 & 2.7 & 0.13 \\ 
2018 Dec 21-22 & 29 - 36 & 102 & 0.48  & 0.27 & 1.6 & 0.25 \\ 
2018 Dec 22-23 & 60 - 65 & 84 & 0.45  & 0.25 & 1.5 & 0.25 \\ 
\hline 
\end{tabular} 
\end{center} 
\caption{Echo bandwidths and disk-integrated properties derived from integrated spectrum for each track. $\Delta \phi$  is the rotation phase coverage observed with each setup, where we have adopted a rotation period of 285 hours (Warner 2016) and the start of reception at 
23:49:12
on 2018 December 20 defines the zero-phase epoch. 
Note that this ignores any apparent rotation due to sky motion.
OC SNR value is obtained from the peak of the spectrum. 
Bandwidth $B$ is measured from the zero-sigma crossings, with uncertainty of 0.02 Hz. 
$\sigma_{OC}$ is the OC radar cross-section and SC/OC is the circular polarization ration estimated from the echo power spectra (Figure 1). $D_{max}$ is the lower bound on the asteroid's maximum pole-on breadth derived from Equation (3) with cos $\delta$ = 
1, which means an equatorial view.
 }

\label{table: 2} 
\end{table}

The echo bandwidths vary from day-to-day by a factor of about 1.8 (Fig. 1; Table 2), that, taken at face value, suggest an elongation of at least 1.8.  This is consistent with the elongation visible in the Goldstone and Arecibo delay-Doppler 
images (Rivera-Valentin et al. 2019). 
If we adopt a rotation period of 285 h, and ignore possible non-principal axis rotation, then the observing intervals spanned 5-7 degrees of rotation on each day and at least several tens of degrees between Dec. 20-23.
 
The echo shapes provide information on the shape of the asteroid. The echoes on Dec. 20-21 (Fig. 1 and 2) and 22-23 (Fig. 1 and 4) are asymmetric and indicate that the asteroid's shape is asymmetric as well.  The echoes on Dec. 22-23 (Figs. 1 and 4) have a conspicuous and statistically significant dip near the middle that resembles dips seen in dozens of other near-Earth asteroids observed by radar.  This probably corresponds to a  large concavity. 
Such a feature is indicative of topography or a more complex shape (i.e., not a simple ellipsoid.
Comparing the timing of the ATCA data on that date with images obtained at Arecibo earlier  on the same date, a large concavity visible in the Arecibo images should have been at least partially oriented toward  the Earth during the Canberra/ATCA observations
, which suggests it could be the concavity/shark-bite feature from the radar images.

A 285 hour rotation period of 2003 SD220 (Warner 2016) means a phase change rate of 1.28 degrees/hour or 30.7 degrees/day. 
Note that this ignores any apparent rotation due to sky motion but this may not be insignificant.
Although the bandwidths and echo shapes vary by only a few degrees within each track, as seen in Figures 2-4, they change significantly from day-to-day.  If we adopt the rotation period of 285 h and again assume principal axis rotation, then the bandwidths listed in Table 2 place lower bounds on the pole-on breadths of 2.7-1.5
km)/cos $\delta$, where delta is the unknown subradar latitude.  These constraints on the dimensions of the object are consistent with visible range extents observed in Arecibo-GBT and Goldstone-GBT delay-Doppler images.  In more detail, the range-Doppler radar images taken 18, 19, and 22 December with the Arecibo-GBT bistatic radar at S-band revealed an elongated shape with a long axis of $>$ 2.5 km and an intermediate axis of roughly 1 km rotating nearly 120 deg over 
four days (Rivera-Valentin et al. 2019).  The Canberra-ATCA echo bandwidths (when converted to a wavelength of 3.5 cm), spectral shapes, and circular polarization ratios are also consistent with the Goldstone-Svetloe and Goldsone-Zelenchukskaya CW observations reported by Bondarenko et al. (2019).  The dimensions suggested by the echo power spectra are 
several times
than the diameter of 0.8 $\pm$ 0.2 km reported by Nugent et al. (2016) from NEOWISE spacecraft observations.  This apparent discrepancy is probably due to Nugent et al.'s adoption of a simple 
spherical
model, which is very different from the highly elongated and irregular shape of the asteroid revealed by radar data.  \

\begin{figure}[!ht] 
\centering 
\includegraphics[width=0.88 \textwidth]{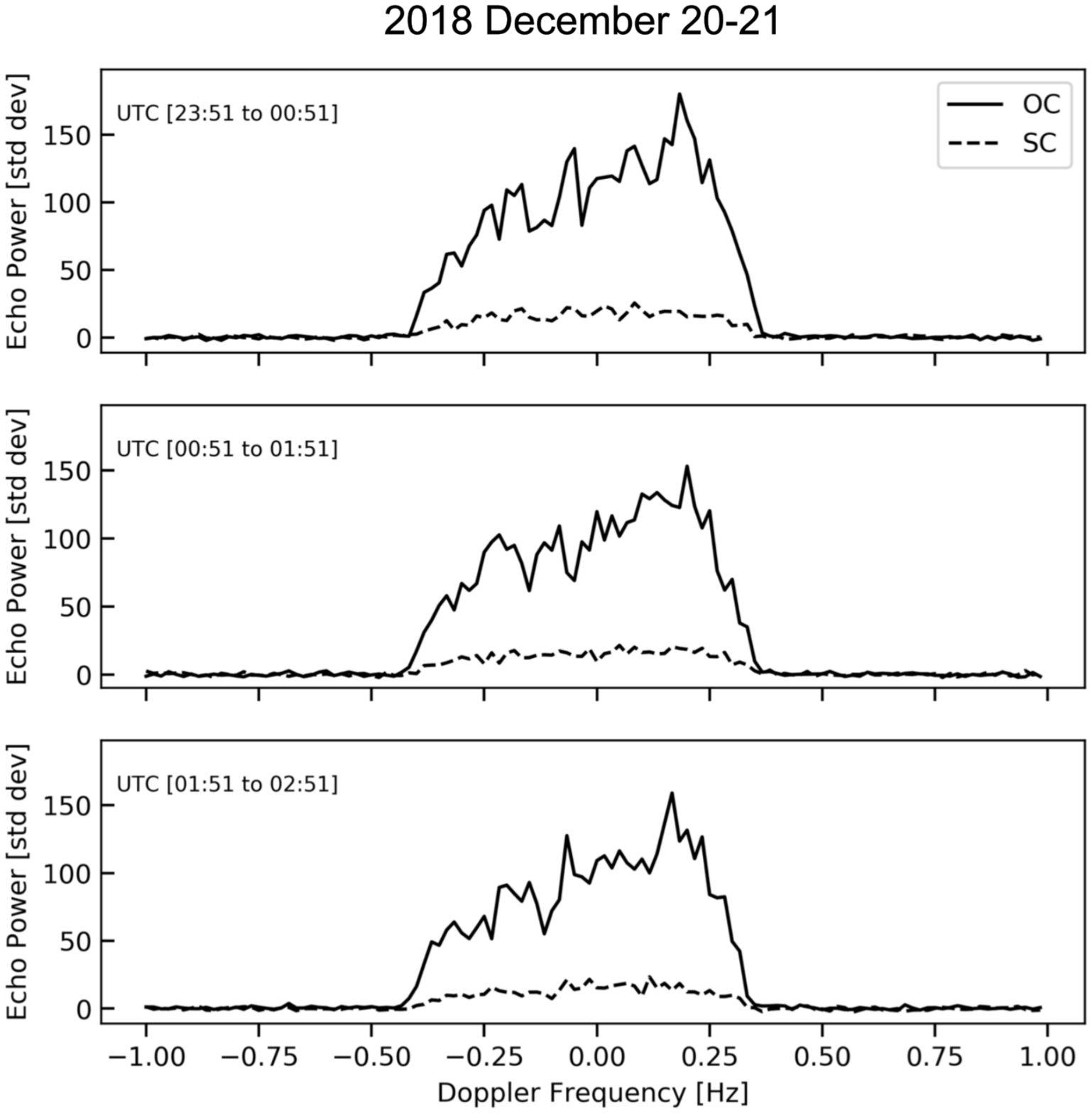} 
\caption{Echo power spectra averaged in 60 minute intervals on Dec. 20-21 (transmitted by DSS-43) at a resolution of 0.0167 Hz. 
The 0 point
of the x-axis in this figure and in the other figures is 7159.45 MHz.}
\label{fig: correction2} 
\end{figure}

\begin{figure}[!ht] 
\centering 
\includegraphics[width=0.88 \textwidth]{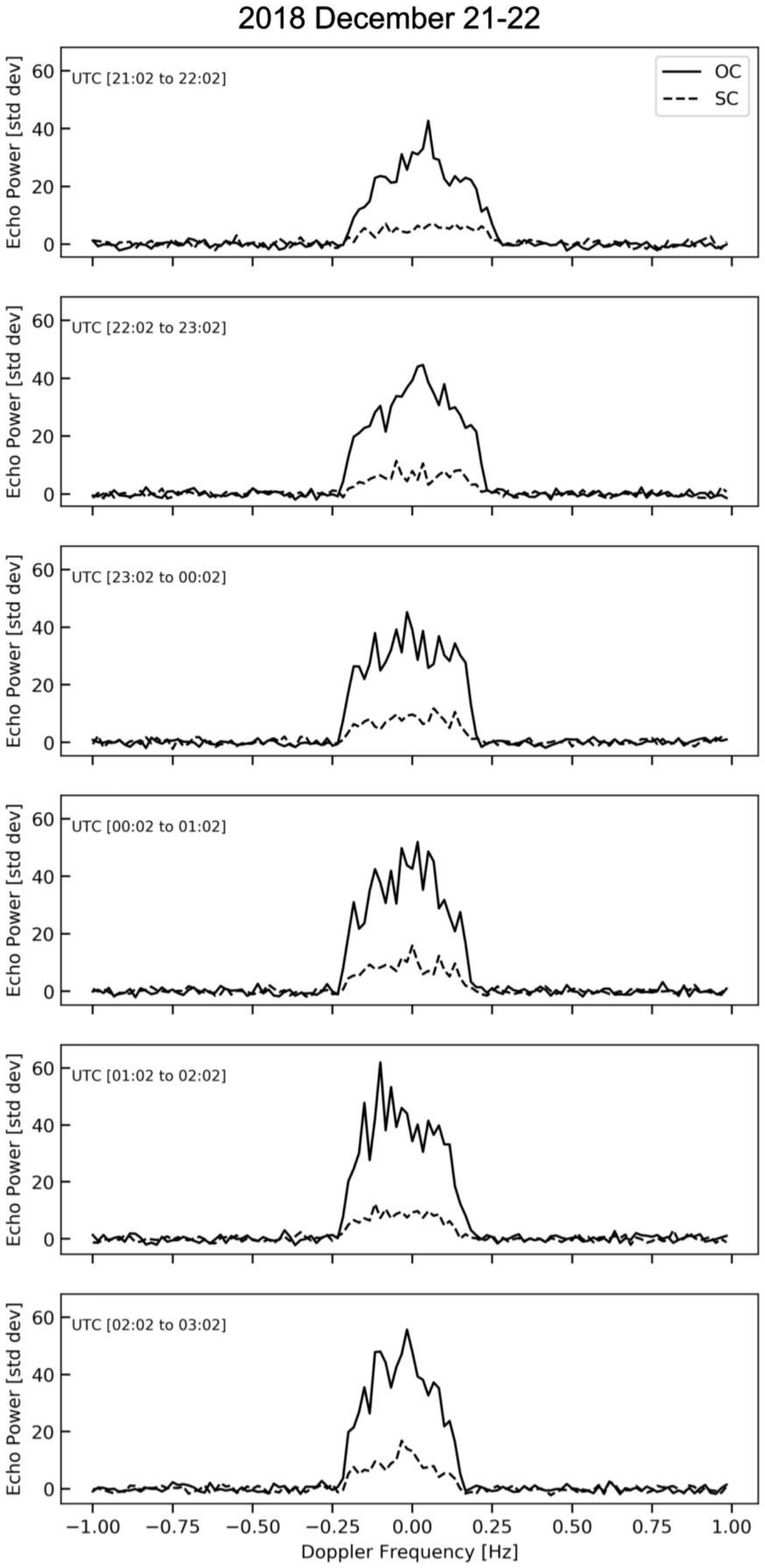} 
\caption{Echo power spectra averaged in 60 minute intervals on Dec. 21-22 (transmitted by DSS-35) at a resolution of 0.0167 Hz.
Despite the very slow rotation period, there appears to be evidence of rotation over time with the shape becoming more end on (narrower).
} 
\label{fig: correction2} 
\end{figure}

\begin{figure}[!ht] 
\centering 
\includegraphics[width=0.88 \textwidth]{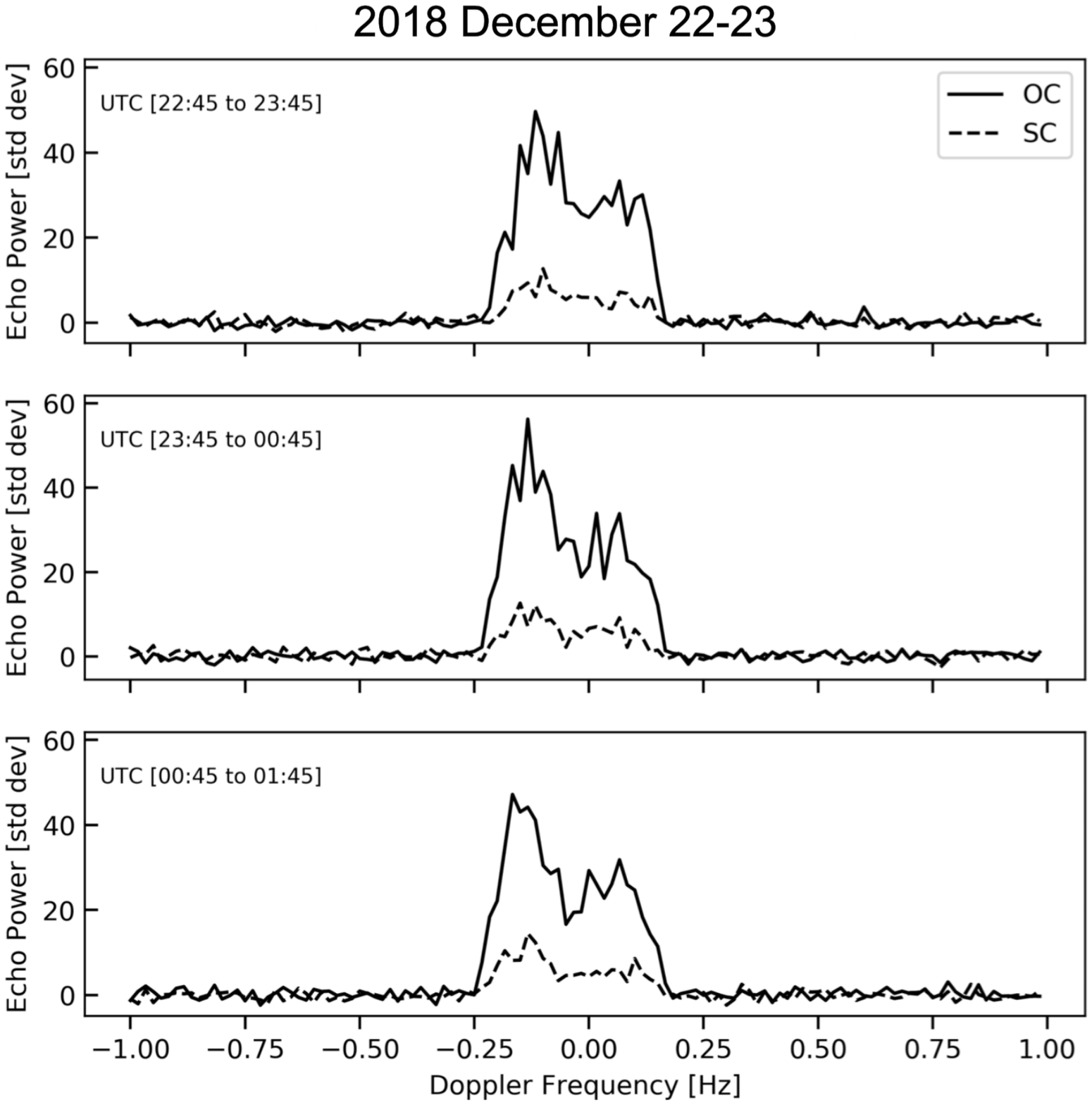} 
\caption{Echo power spectra averaged in 60 minute intervals on Dec. 22-23 (transmitted by DSS-36) at a resolution of 0.0167 Hz.
The dip in the middle of the echo may be indicative of topography
} 
\label{fig: correction2} 
\end{figure}

 Weighted sums of all the data from each day yield OC radar cross sections of 0.39, 0.27, and 0.25 km$^{2}$ $\pm$ 35 \% that are consistent with preliminary results obtained at Arecibo. 
 We obtain daily SC/OC ratios between 0.13 $\sim$ 0.25 with a mean and rms dispersion of 0.21 $\pm$ 0.07; variations of this magnitude are not unusual for near-Earth asteroids (Benner et al. 2008).  The uncertainty in the SC/OC ratio for each day is small, estimated to be 0.02 as a typical error, considering the circular polarizations at the ATCA are digitally converted in the correlator from linear polarization feeds. 
 The small uncertainty on SC/OC is because any gain, calibration, pointing errors should affect both polarizations equally, so the margin given to the peak measurement does not apply to the quotient of the SC and OC peaks.
The daily variations
may represent real variations in the radar scattering properties of this asteroid as seen at different orientations, which changed markedly from day to day.
Detailed analysis of extensive Goldstone and Arecibo observations may help understand these results.

\section{Discussion} \label{Section4}

Benner et al. (2008) compiled circular polarization ratios for 214 NEAs that revealed a strong correlation between SC/OC and some spectral classes.   Benner et al. found that the C-class taxonomy ($N$ = 17) has 
S/C = 0.28
$\pm$ 0.12, the S class has SC/OC 0.27 $\pm$ 0.08 ($N$ = 70), and the E, V, and X taxonomies have a mean SC/OC $>$ 0.6. Our mean circular polarization ratio of SC/OC = 0.21 $\pm$ 0.07 (Table 2) is consistent with the S-class results reported 
for other NEAs by Benner et al. (2008) and also with the S-type taxonomy from infrared spectroscopy for 2003 SD220 reported by Rivera-Valentin et al. (2019).
In Table 3 we compile SC/OC values estimated by radar experiments for  near-Earth asteroids visited by spacecraft, where there's some ground-truth about the roughness of the near-surface.  Within their stated uncertainties, the average ratio we obtain for 2003 SD220 is consistent with those reported for all other NEAs observed by radar that have also been visited by spacecraft regardless, although SC/OC for 2003 SD220, taken at face value, is slightly less than values obtained for Eros, Itokawa, and Toutatis.  
Rivera-Valentin et al. (2019) report CW experiments during the 2015 and 2018 apparition revealed a SC/OC of  $\sim$ 0.2.
Bondarenko et al. (2019) estimate SC/OC of 0.22 and 0.25 for two epochs around the same time of our observations. 
Our result is consistent with those measurements. 

\begin{table}[!h] 
\begin{center} 
\small 
\setlength{\tabcolsep}{7pt} 
\begin{tabular}{lccccc} 
\hline 
Asteroids  & Class & Mission &  SC/OC & Band & Reference \\ 
\hline 
433 Eros & S & NEAR-Shoemaker  &  0.28 $\pm$0.06 & S & Magri et al. 2001 \\ 
25143 Itokawa  & S & Hayabusa & 0.26$\pm$0.04  & S & Ostro et al. 2004 \\ 
4179 Toutatis & S & Chang'e 2 &  0.29$\pm$0.01  & X & Ostro et al. 1999 \\ 
2003 SD220  & S & -  & 0.21$\pm$0.07  & C & This work  \\ 
101955 Bennu & C & OSIRIS-REx &  0.19$\pm$0.03 & X & Nolan et al. 2013 \\ 
101955 Bennu & C & OSIRIS-REx &  0.18$\pm$0.03 & S & Nolan et al. 2013 \\ 
162173 Ryugu & C & Hayabusa2 &  -  & - & -  \\ 
\hline 
\end{tabular} 
\end{center} 
\caption{The SC/OC values estimated by radar experiments for  near-Earth asteroids visited by spacecraft. Arecibo observed at S-band and Goldstone DSS-14 observed at X-band.  Ryugu is a radar target at Arecibo and Goldstone in December, 2020.
Note that relative comparison of SC/OC of different taxonomies may not translate directly to relative surface roughness.} 
\label{table: 2} 
\end{table}

The next opportunity for radar observations of 2003 SD220 will be in late 2021, at 0.0363 au (14.2 lunar distances), on 2021 December 17. 
The distance will be twice as far from the Earth than in 2018 so the SNRs will be 16 times lower if all else is equal.
But by early 2021 an 80 kW transmitter will be operational at DSS-43, so we could obtain SNRs only a factor of 4 less than the results presented here,
which will be valuable for 3-D modelling and spin state estimation.
In addition, we plan to utilise a wave form generator for delay-Doppler analysis and imaging for our southern hemisphere asteroid radar project by then.

\section*{Acknowledgments}

We wish to thank the staff of Canberra Deep Space Communication Complex (CDSCC) for their assistance. BM was supported by the CSIRO Vacation Student Scholarship while part of this research was conducted. Part of this research was carried out at the Jet Propulsion Laboratory, California Institute of Technology, under a contract with the National Aeronautics and Space Administration.

\section*{References}

\bibliographystyle{elsarticle-harv}

\bibliography{references}

Abu-Shaban, Z. et al., 2018, Asteroids Observation from the Southern Hemisphere Using Planetary Radar. 42nd COSPAR Scientific Assembly. Held 14-22 July 2018, in Pasadena, California, USA, Abstract id. B1.1-79-18.
ICARUS
Benson, C. et al., 2017a, First Detection of Two Near-Earth Asteroids With a Southern Hemisphere Planetary Radar System. Radio Science, 52, 1344-1351.

Benson, C. et al., 2017b, Detection of three near-earth asteroids with a southern hemisphere planetary radar system. in Proceedings of the International Astronautical Congress, IAC, pp. 2959 - 2964

Benner, L. A. M. et al. 2008, Near-Earth asteroid surface roughness depends on compositional class. ICARUS, Vol. 198, pp. 294-304.

Bondarenko,Y. S., Marshalov, D. A., Vavilov, D. E., Medvedev, Y. D. 2019, Radar observations of near-Earth asteroid 2003 SD220. EPSC-DPS Joint Meeting 2019. 

DeMeo, F. E., Binzel, R. P., Lockhart, M. 2014.  Mars encounters cause fresh surfaces on some near-Earth asteroids.  ICARUS 227, 112-122. 

Kawakatsu, Y., Abe, M., Kawaguchi, J., 2009, Trans. JSASS Space Tech. Japan, 3, 33-41. 

Gaffey, M., Gilbert, S., 1998, Meteorit. Planet. Sci. 33, 1281-1295. 

Ostro, S.J., 1993, Planetary Radar Astronomy. Rev. of Mod. Phys., 65, 1235-1279.

Ostro, S.J. et al. 1999, Asteroid 4179 Toutatis: 1996 radar observations.  ICARUS 137, 122-139.

Ostro, S.J. et al. 2004, Radar observations of asteroid 25143 Itokawa (1998 SF36).  Meteorit. Planet.  Sci. 39, 407-424.  

Magri et al. 2001, Radar constraints on asteroid regolith compositions using 433 Eros as ground truth.  Meteorit. Planet. Sci. 36, 1697-1709.

Nolan, M. C. et al. 2013, Shape model and surface properties of the OSIRIS-REx target asteroid (101955) Bennu  from radar and lightcurve observations.  ICARUS 226, 629-640.

Nugent, C. R. et al. 2016, NEOWISE Reactivation Mission Year Two: Asteroid Diameters and Albedos, Astron.l J., 152, 63.

Perna, D. et al. 2016, Grasping the nature of potentially hazardous asteroids.  Astron. J. 151, 11.

Phillips, C. J. et al, 2009, The LBADR: the LBA Data Recorder, 8th International e-VLBI Workshop - EXPReS09, Madrid, Spain, 99p 

Rivera-Valentin, E. G., et al. 2019, Radar and near-infrared characterization of near-earth-asteroid (163899) 2003 SD220. 50th Lunar and Planetary Science Conference 2019 (LPI Contrib. No. 2132).

Warner, B. D., 2016, Near-Earth asteroid lightcurve analysis at CS3-Palmer Divide Station: 2015 October-December. Minor Planet Bull., Vol. 43, pp. 143-154.

\end{document}